\documentclass[sap,reprint,showpacs,superscriptaddress,groupedaddress,balancelastpage,floatfix]{revtex4-1}  

\usepackage[utf8]{inputenc}
\usepackage{graphicx,dcolumn,bm,amssymb,amsmath,amsfonts,xcolor}
\usepackage[
pdfstartview=FitBV,
bookmarks=true,
bookmarksopen=true,
colorlinks =true,
linkbordercolor=blue,
linkcolor = blue,
citecolor = blue,
urlcolor=blue]{hyperref}
\usepackage[capitalize]{cleveref} 

\crefname{section}{Sec.}{Sections}
\crefname{table}{TABLE}{TABLEs.}
\crefname{figure}{FIG.}{FIGs.}

\begin{document}
	
	\title{On the formation of van der Waals complexes through three-body recombination}
	\author{Marjan Mirahmadi}
	\email{m.mirahmadi@fhi-berlin.mpg.de}
	\affiliation{Fritz-Haber-Institut der Max-Planck-Gesellschaft, Faradayweg 4-6, D-14195 Berlin, Germany}
	\author{Jes\'us P\'{e}rez-R\'{i}os}
	\email{jperezri@fhi-berlin.mpg.de}
	\affiliation{Fritz-Haber-Institut der Max-Planck-Gesellschaft, Faradayweg 4-6, D-14195 Berlin, Germany}
	
	\begin{abstract}
		In this work, we show that van der Waals molecules X-RG (where RG is the rare gas atom) may be created through direct three-body recombination collisions, i.e., X + RG + RG $\rightarrow$ X-RG + RG. In particular, the three-body recombination rate at temperatures relevant for buffer gas cell experiments is calculated via a classical trajectory method in hyperspherical coordinates [J. Chem. Phys. \textbf{140}, 044307 (2014)]. As a result, it is found that the formation of van der Waals molecules in buffer gas cells (1~K $\lesssim T \lesssim 10$ K) is dominated by the long-range tail (distances larger than the LeRoy radius) of the X-RG interaction. For higher temperatures, the short-range region of the potential becomes more significant. Moreover, we notice that the rate of formation of van der Walls molecules is of the same order of magnitude independently of the chemical properties of X. As a consequence, almost any X-RG molecule may be created and observed in a buffer gas cell under proper conditions. 
		
	\end{abstract}
	
	\maketitle
	
	\section{Introduction}\label{sec:intro}
	When a three-body process leads to the formation of a molecule as a product state, A + A + A $\rightarrow$ A$_2$ + A, it is labelled as a three-body recombination process or as a ternary association reaction. These three-body processes are relevant for a wide variety of systems in areas ranging from astrophysics to ultracold physics. In particular, three-body recombination of hydrogen is one of the essential processes to explain H$_2$ formation in star-forming regions \cite{Flower2007,Forrey}. In the field of ultracold physics, recent developments in laser technologies and cooling techniques have made it possible to gain a more in-depth insight into the significant role of three-body recombination processes in different phenomena such as atomic loss processes in ultracold dilute gases \cite{Esry1999,Weiner1999,Bedaque2000,Suno2003,Weber2003,Schmidt2020,Greene2017}, and the formation and trapping of cold and ultracold molecules \cite{Koehler2006,Blume2012,Perez-Rios2015,Krukow2016,mohammadi2020life}.
	
	Van der Waals (vdW) molecules consist of two atoms held together by the long-range dispersion interaction \cite{Blaney1976} presenting binding energies $\lesssim 1$ meV. Therefore, vdW molecules show the weakest gas-phase molecular bond in nature~\footnote{Except for ultra-long-range Rydberg molecules showing binding energies $\sim 4$ neV \cite{Greene2000,Fabikant,Hamilton2002,Pfau,Trilobite,Butterfly}.}. The binding mechanism in vdW molecules relies on the compensation between the short-range repulsive interaction (due to the overlap of closed-shell orbitals) and the attractive  $-C_6/r^6$ vdW interaction, where the dispersion coefficient $C_6$ depends on the polarizability of the interacting atoms. Interestingly enough, the study of vdW interactions provides crucial information necessary to investigate the formation and stability of gases, liquids, and materials such as vdW heterostructures and biopolymers \cite{Buckingham1988,Koperski2002,Levine2005,Hermann2017}; chemical reactions \cite{Smalley1977,Worsnop1986,Skouteris1999,Balakrishnan2004,Levine2005,Shen2017}; and physical phenomena like superfluidity of $^{4}$He nanodroplets \cite{Toennies2004,Szalewicz2008}. In particular, investigating the properties of vdW molecules (as the simplest form of vdW complexes) containing a rare gas atom leads to a deeper understanding of the nature of bonding in rare gas crystals, and of the dynamics of impurities interacting with dense rare gas vapors \cite{Buckingham1938,Tangt1976,Fugol1978,Brahms2011}. 
	
	Despite the significance of vdW molecules in modern chemical physics, the community has been focused on its characterization rather than on revealing how they emerge in different scenarios \cite{Smalley1977,Tellinghuisen1979,Worsnop1986,Martrenchard1993,Koperski2002,Brahms2008,Brahms2010,Brahms2011,Tariq2013,Friedrich2013,Quiros2017}. Recently, thanks to the development of buffer gas sources \cite{DeCarvalho1999}, it has been possible to investigate the formation of vdW molecules through three-body recombination processes \cite{Brahms2008,Suno2009,Brahms2010,Brahms2011,Wang2011,Tariq2013,Quiros2017}. However, the field is still lacking a global study on the formation of vdW molecules through three-body collisions.
	
	 In the present work, we study the formation of vdW molecules X-RG  through direct three-body recombination processes X + RG + RG $\rightarrow$ X-RG + RG. Here RG indicates the rare gas atom and atom X has been chosen in a way to cover a broad range of chemical characteristics, i.e., from three different groups of the periodic table: alkali group (Li and Na), transition metals (Ti), and pnictogen group (As, P, and N). Our approach is based on a classical trajectory (CT) methodology in hyperspherical coordinates, which has been already applied to three-body recombination of helium \cite{Perez-Rios2014,Greene2017}, and to ion-neutral-neutral three-body recombination processes \cite{Perez-Rios2015,Krukow2016,Perez-Rios2018}. Indeed, a direct three-body approach for forming vdW molecules has never been carried out via CT method to the best of our knowledge. Performing these calculations, we notice a clear distinction between the formation rates at low-energy collisions and high-energy collisions, established by the dissociation energy of the X-RG potential. Moreover, our results show that the three-body recombination rate is of the same order of magnitude independently of the X atom, and hence most of the vdW molecules X-RG should be observable in buffer gas cells.  
	
	This paper is organized as follows: In \cref{sec:CT}, we summarise the main aspects of the classical trajectory method employed to study direct three-body recombination processes. In \cref{sec:results}, we precisely investigate the dependence of three-body recombination rates on the collision energy and temperature, utilizing six different systems. In \cref{sec:rel}, we discuss the applicability of the classical treatment at low temperatures. Finally, in \cref{sec:conclusion}, we summarize our chief results and discuss their possible applications. 
	
	\section{Classical trajectory method in hyperspherical coordinates}\label{sec:CT}
		\begin{figure}[b]
		\begin{center}
			\includegraphics[scale=0.6]{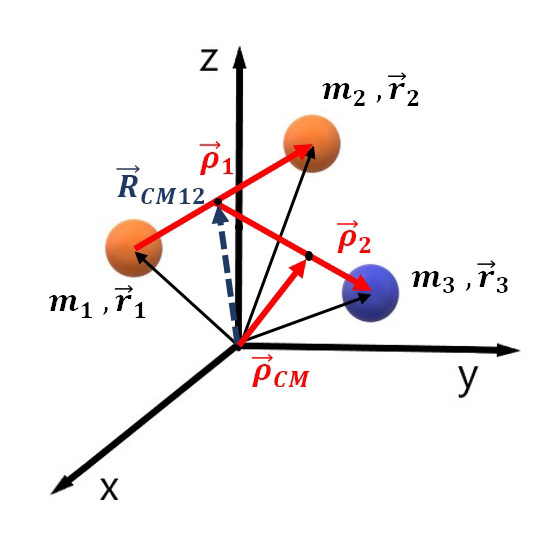}
			\caption{\label{fig:jacobi}Jacobi coordinates for the three-body problem. Here, $\vec{R}_{cm12}$ is the center-of-mass vector of two-body system consists of $m_1$ and $m_2$. }
		\end{center}
	\end{figure} 
	
	Consider a system of three particles with masses $m_i$ ($i = 1,2,3$) at the respective positions $\vec{r}_i$, interacting with each other via the potential $V(\vec{r}_1,\vec{r}_2,\vec{r}_3)$. Here, we neglect the three-body term of the potential,  which is a good approximation for van der Waals molecules and clusters \cite{3bodyinteractions}, hence $V$ can be expressed as a summation of pair-wise potentials, i.e., $	V(\vec{r}_1,\vec{r}_2,\vec{r}_3) = U(r_{12}) +  U(r_{23}) +  U(r_{31})$, where $r_{ij} = |\vec{r}_j - \vec{r}_i|$. The dynamics of these particles is governed by the Hamiltonian 
	
	\begin{equation}\label{eq:cartesianH}
		H = \frac{\vec{p}_1^{~2}}{2m_1} + \frac{\vec{p}_2^{~2}}{2m_1} + \frac{\vec{p}_3^{~2}}{2m_1} + U(r_{12}) +  U(r_{23}) +  U(r_{31}) \quad
	\end{equation}
	with  $\vec{p}_i$ being the momentum vector of the $i$-th particle.
	
	To solve Hamilton's equations and find classical trajectories, it is more convenient to employ Jacobi coordinates \cite{Pollard1976,suzuki1998}. For a three-body problem, Jacobi vectors are related to $\vec{r}_i$ vectors as 
	
	\begin{align}\label{jacobitrans}
		\vec{\rho}_1 &= \vec{r}_2 - \vec{r}_1 ~, \nonumber \\
		\vec{\rho}_2 &= \vec{r}_3 - \frac{m_1\vec{r}_1 + m_2\vec{r}_2}{m_1+m_2} ~, \nonumber \\
		\vec{\rho}_{CM} &= \frac{m_1\vec{r}_1 + m_2\vec{r}_2 + m_3\vec{r}_3}{M} ~,
	\end{align}
	where $M = m_1 + m_2 + m_3$ is the total mass of the system and $\vec{\rho}_{CM}$ is the three-body center-of-mass vector. These vectors are illustrated in \cref{fig:jacobi}.

	Due to the conservation of total linear momentum (i.e., $\vec{\rho}_{CM}$ is a cyclic coordinate), the degrees of freedom of the center of mass can be neglected, thus, the Hamiltonian~\eqref{eq:cartesianH} transforms to
	\begin{equation}\label{eq:jacobiH}
		H = \frac{\vec{P}_1^2}{2\mu_{12}} + \frac{\vec{P}_2^2}{2\mu_{3,12}} +  V(\vec{\rho}_1,\vec{\rho}_2) ~.
	\end{equation} 
	Here, $\mu_{12}=m_1m_2/(m_1 + m_2)$ and $\mu_{3,12}=m_3(m_1+m_2)/M$; and  $\vec{P}_1$ and $\vec{P}_2$ indicate the conjugated momenta of $\vec{\rho}_1$ and $\vec{\rho}_2$, respectively. $V(\vec{\rho}_1,\vec{\rho}_2) $ is the potential expressed in terms of the Jacobi coordinates. 
	
	Noting that the Hamilton's equations of motion are invariant under the canonical transformation~\eqref{jacobitrans}, it is possible to predict the evolution of the trajectories in terms of Jacobi coordinates from Hamiltonian~\eqref{eq:jacobiH} via
	
	\begin{align}\label{eq:Heqns}
		\frac{d\vec{\rho}_i}{d t} = \frac{\partial H}{\partial \vec{P}_i} ~, \quad \quad 
		\frac{d\vec{P}_i}{d t} = -\frac{\partial H}{\partial \vec{\rho}_i} ~,
	\end{align}
	\begin{figure}[t]
		\begin{center}
			\includegraphics[scale=0.5]{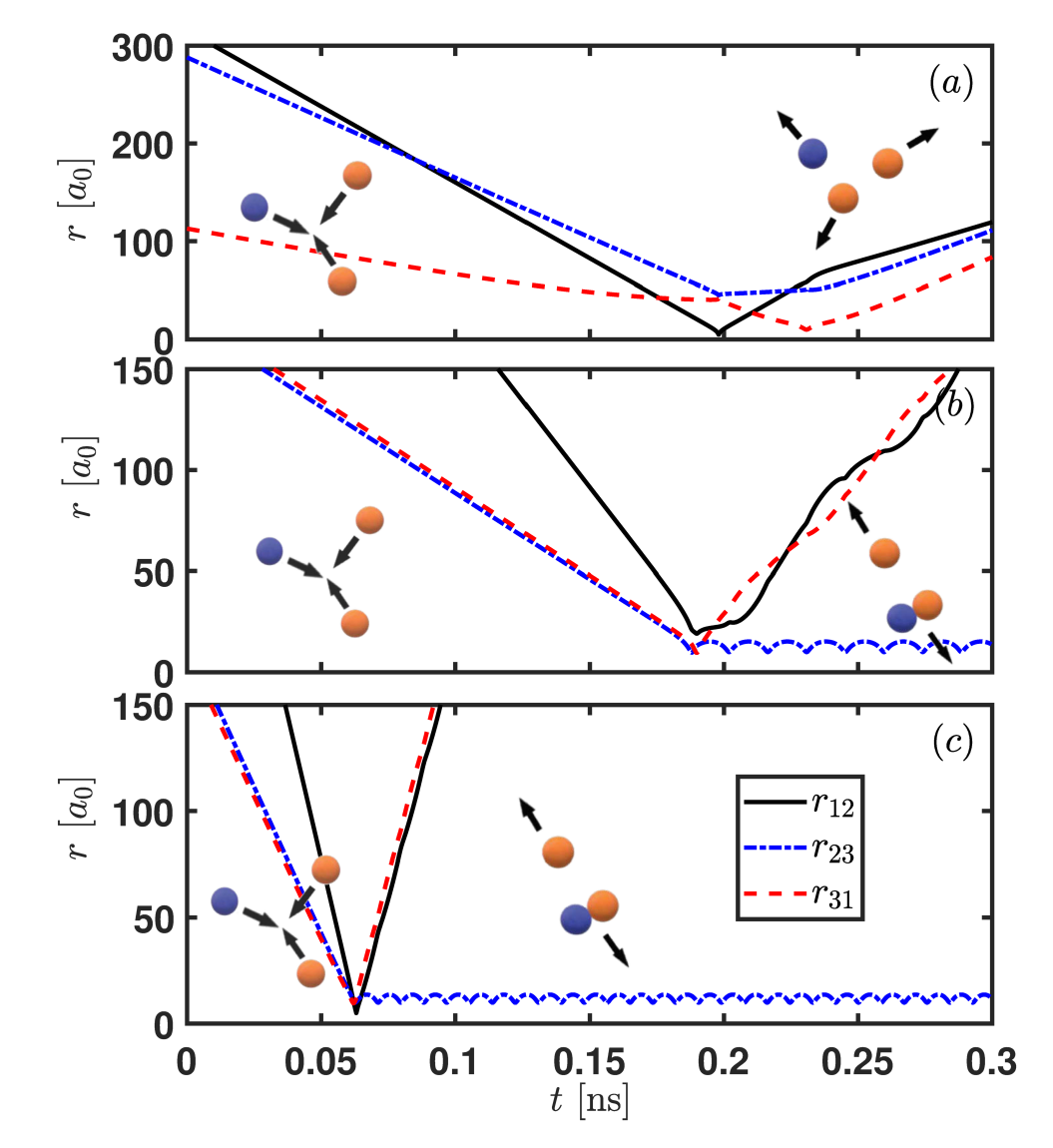}
			\caption{\label{fig:cltraj}Classical trajectories of  Li + He + He $(a)$ elastic collision, $(b)$ recombination event at $E_c = 1$ K with $b = 0$, and $(c)$ recombination event at $E_c = 10$ K with $b = 0$.}
		\end{center}
	\end{figure}
	and transform the solutions back to Cartesian coordinates. 
	As an example, \cref{fig:cltraj} shows the classical trajectories calculated for Li + He + He three-body collisions for different collision energies and the same impact parameter ($b=0$). The panel $(a)$ of this figure shows an elastic or non-reactive trajectory in which the three-body collision leads to three free particles flying away form each other. On the contrary, in panels $(b)$ and $(c)$, the three-body collision ends up forming a molecule that vibrates rapidly, i.e, a three-body recombination event Li + He + He $\rightarrow$ Li-He + He. 

	\subsection{Classical three-body recombination in hyperspherical coordinates}\label{subsec:CTBR}
	It is well-known that, classically, $n$-body collisions in a three-dimensional (3D) space can be mapped into a problem involving one particle with a definite momentum moving towards a scattering center in a $d$-dimensional space in which $d=3n-3$ is equal to the independent relative coordinates of the $n$-body system. Exploiting this point, we define the initial conditions and impact parameter associated with a three-body problem as single entities in a six-dimensional (6D) space. The 6D space is described in hyperspherical coordinates,  which consist of a hyperradius $R$, and five hyperangles $\alpha_j$ ($j = 1,2,3,4,5$), where $0\leq\alpha_1<2\pi$ and $0\leq\alpha_{j>1}\leq\pi$ \cite{Smith1960,Smith1962,Whitten1968,Johnson1980,Johnson1983,Lin1995,Avery2012}. 
	
	Position and momentum vectors in the hyperspherical coordinates can be constructed from Jacobi vectors and their conjugated momenta as 
	
	\begin{align}\label{eq:rho6D}
		\vec{\rho} = \begin{pmatrix} \vec{\rho}_1 \\ \vec{\rho}_2 \end{pmatrix} ~,
	\end{align}
	and 
	
	\begin{align}\label{eq:P6D}
		\vec{P} = \begin{pmatrix} \sqrt{\frac{\mu}{\mu_{12}}}\vec{P}_1 \\ \sqrt{\frac{\mu}{\mu_{3,12}}}\vec{P}_2 \end{pmatrix} ~,
	\end{align}
	respectively, where $\mu = \sqrt{m_1 m_2 m_3/ M}$ is the three-body reduced mass (for further details see Refs.~[\onlinecite{Perez-Rios2014,Perez-Rios2020}]). Consequently, the Hamiltonian $H$ in this coordinates reads as
	
	\begin{equation}\label{eq:6DH}
		H = \dfrac{\vec{P}^2}{2\mu} + V(\vec{\rho}) ~.
	\end{equation}
	
	In the 3D space, the collision cross section $\sigma$ is defined as the area drawn in a plane perpendicular to the initial momentum containing the scattering center, that the relative motion of the particles (known as trajectory) should cross in order to a collision to take place \cite{Levine2005}. This concept can be extended to a 6D space by visualizing it in a five-dimensional hyperplane (embedded in a 6D space) instead of a plane \cite{Perez-Rios2014,Perez-Rios2020}.
	Using the same analogy, we can define the impact parameter vector $\vec{b}$ as projection of the position vector in a 5D hyperplane perpendicular to the initial 6D momentum vector $\vec{P}_0$ (i.e., $\vec{b}.\vec{P}_0 = 0$). Therefore, the cross section associated with the three-body recombination process, after averaging over different orientations of $\vec{P}_0$, is obtained as follows \cite{Perez-Rios2014}:
	
	\begin{align}\label{eq:sigma}
		\sigma_{rec}(E_c) & = \int \mathcal{P}(E_c,\vec{b})  b^4 d b ~d\Omega_b \nonumber \\
		&= \frac{8\pi^2}{3}\int_{0}^{b_{max}(E_c)} \mathcal{P}(E_c,b)  b^4 d b ~,
	\end{align}
	where $d\Omega_b = \sin^3(\alpha_4^b)\sin^2(\alpha_3^b)\sin(\alpha_2^b)d\alpha_4^b d\alpha_3^b d\alpha_2^b d\alpha_1^b$ is the solid angle element associated with vector $\vec{b}$, and  we made use of the relation $P_0 = \sqrt{2\mu E_c}$. The function $\mathcal{P}$ is the so-called opacity function, i.e., the probability that a trajectory with particular initial conditions leads to a recombination event. Note that the factor $\Omega_b = 8\pi^2/3$ is the solid hyperangle associated with $\vec{b}$, and $\mathcal{P}(E_c,b) = 0$ for $b>b_{max}$. In other words,   $b_{max}$ represents the largest impact parameter for which three-body recombination occurs. Finally, the energy-dependent three-body recombination rate is obtained as
	
	\begin{equation}\label{eq:k3}
		k_3(E_c) = \sqrt{\frac{2E_c}{\mu}}\sigma_{rec}(E_c) ~.
	\end{equation}
	
	\subsection{Computational details}\label{subsec:comput}
	The angular dependence of the opacity function $\mathcal{P}(\vec{b},\vec{P}_0)$, which depends on both direction and magnitude of impact parameter and initial momentum vectors, has been averaged out by means of Monte Carlo method \cite{Landau2014,Perez-Rios2014}. Without loss of generality, we choose the $z$ axis in 3D space to be parallel to the Jacobi momentum vector $\vec{P}_2$. The initial hyperangles determining the orientation of vectors $\vec{P}_0$ and $\vec{b}$ in the 6D space are sampled randomly from probability distribution functions associated with the appropriate angular elements in hyperspherical coordinates (see Ref.~[\onlinecite{Perez-Rios2014}]).
	
	In the next step, the opacity function $\mathcal{P}(E_c,b)$ for a given collision energy, $E_c = P_0^2/(2\mu)$ and magnitude of impact parameter, $b$, is obtained by dividing the number of classical trajectories that lead to recombination events, $n_r$, by the total number of trajectories simulated $n_t$ \cite{Perez-Rios2014}. Thus,
	
	\begin{align}\label{delta}
		\mathcal{P}(E_c,b) \approx ~ & \frac{n_r(E_c,b)}{n_t(E_c,b)} \pm \nonumber \\
		& \frac{\sqrt{n_r(E_c,b)}}{n_t(E_c,b)}\sqrt{\frac{n_t(E_c,b)-n_r(E_c,b)}{n_t(E_c,b)}} ~,
	\end{align}
	where the second term in \cref{delta} is the statistical error owing the inherent stochastic nature of the Monte Carlo technique. 
	
	To solve the Hamilton's equations we made use of the explicit Runge-Kutta (4,5) method, the Dormand-Prince pair \cite{Dormand1986}. The acceptable error for each time-step has been determined by absolute and relative tolerances equal to $10^{-15}$ and $10^{-13}$, respectively. The total energy is conserved during collisions to at least four significant digits and the magnitude of the total angular momentum vector, $J = |\vec{\rho}_1 \times \vec{P}_1 + \vec{\rho}_2 \times \vec{P}_2|$, is conserved to at least six significant digits. The initial magnitude of hyperradius, $|\vec{\rho}_0|$, is generated randomly from the interval $[R_0-25, R_0+25]  ~ a_0$ centered around $R_0 = 550 ~ a_0$ ($a_0 \approx 5.29 \times 10^{-11} \mathrm{m}$ is the Bohr radius). This value fulfils the condition for three particles to be initially in an uniform rectilinear state of motion.
	
	\begin{figure}[b!]
		\begin{center}
			\includegraphics[scale=0.42]{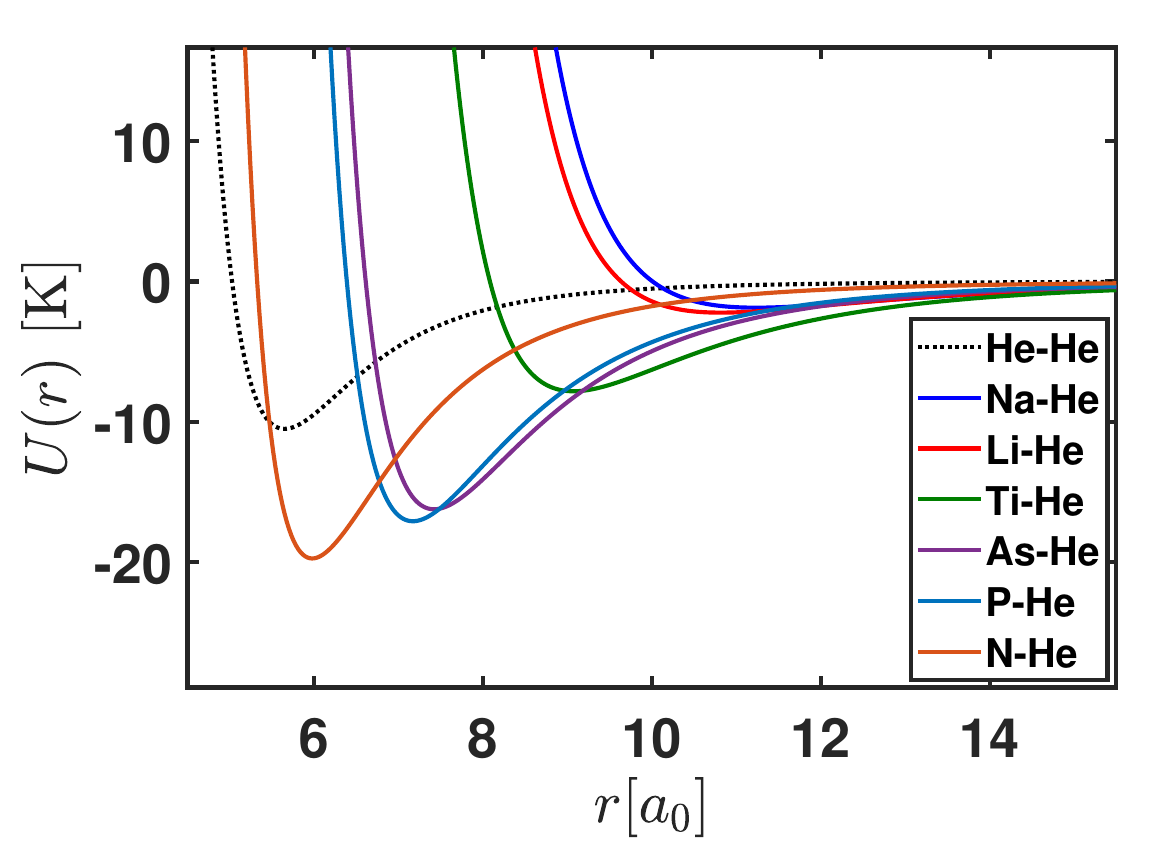}
			\caption{\label{fig:6atompots}X-He potential curves, $U(r)$, of six different atoms. The potentials are obtained from parameters in Refs.~[\onlinecite{Kleinekathofer1999,Tao2010,Tangt1976,Partridge2001,Cambi1991}] (see the text and \cref{tab:table1}). The black dotted curve indicates the He-He interaction based on parameters given in Ref.~[\onlinecite{Aziz1995}].}
		\end{center}
	\end{figure}
		
	\section{Results and discussion}\label{sec:results}	
	Throughout this section, we consider the formation of weakly bound He-containing vdW molecules in their electronic ground state, through the three-body recombination process X + $ ^{4}$He + $ ^{4}$He $\rightarrow$ X-$ ^{4}$He + $ ^{4}$He, for six different X atoms from three different groups of the periodic table. We consider $ ^{7}$Li and $ ^{23}$Na from the alkali group, $ ^{48}$Ti from the transition metals, and  $ ^{75}$As, $ ^{31}$P and $ ^{14}$N from the pnictogen group. All these atoms, with the exception of Ti, show an $S$ electronic ground state.
	
	\Cref{fig:6atompots} displays the two-body potentials $U(r)$ that have been used in the calculations (Li-He and Na-He from Ref.~[\onlinecite{Kleinekathofer1999}]; Ti-He from Refs.~[\onlinecite{Krems2005,Quiros2017}]; As-He, P-He and N-He from Ref.~[\onlinecite{Partridge2001}]; and He-He from Ref.~[\onlinecite{Aziz1995}]). Note that all the X-He complexes show a single electronic state correlated with the ground electronic state of the atom and the rare gas atom, which are described as Lennard-Jones (LJ) potentials with the form $U(r) = C_{12}/r^{12} - C_6/r^6$. However, since the electronic ground state of Ti presents an $F$ symmetry, Ti-He shows four different electronic states correlated with the ground electronic state of Ti and He atoms. In this case, we have taken the LJ potential fitted to the spherically symmetric component of the potential given as
	\cite{Krems2004,Krems2005,Aquilanti}
	
	\begin{equation}\label{TiHepot}
		U(r) = \frac{1}{7} \left[U_\Sigma (r) + 2 U_\Pi (r) + 2 U_\Delta (r) + 2 U_\Phi (r)\right] ~,
	\end{equation}
	where $U_\Sigma$,  $U_\Pi$, $U_\Delta$, and $ U_\Phi$ are the distinct molecular potentials correlated with Ti-He in the ground electronic state. Note that the corresponding well-depths range from $D_e \approx 1.87$ K $\approx 1.30 ~\mathrm{cm}^{-1}$ (for Na-He) to $D_e \approx 19.74$ K $\approx 13.72 ~\mathrm{cm}^{-1}$ (for N-He). 
	
	In what follows, we present the three-body recombination rates calculated from the CT method and explore their dependence on the collision energy and on the particular features of the underlying two-body potentials. 
	
	\subsection{Energy-dependent three-body recombination rate for X-He-He systems }\label{subsec:K_E}

\begin{figure*}
		\begin{center}
			\includegraphics[scale=0.41]{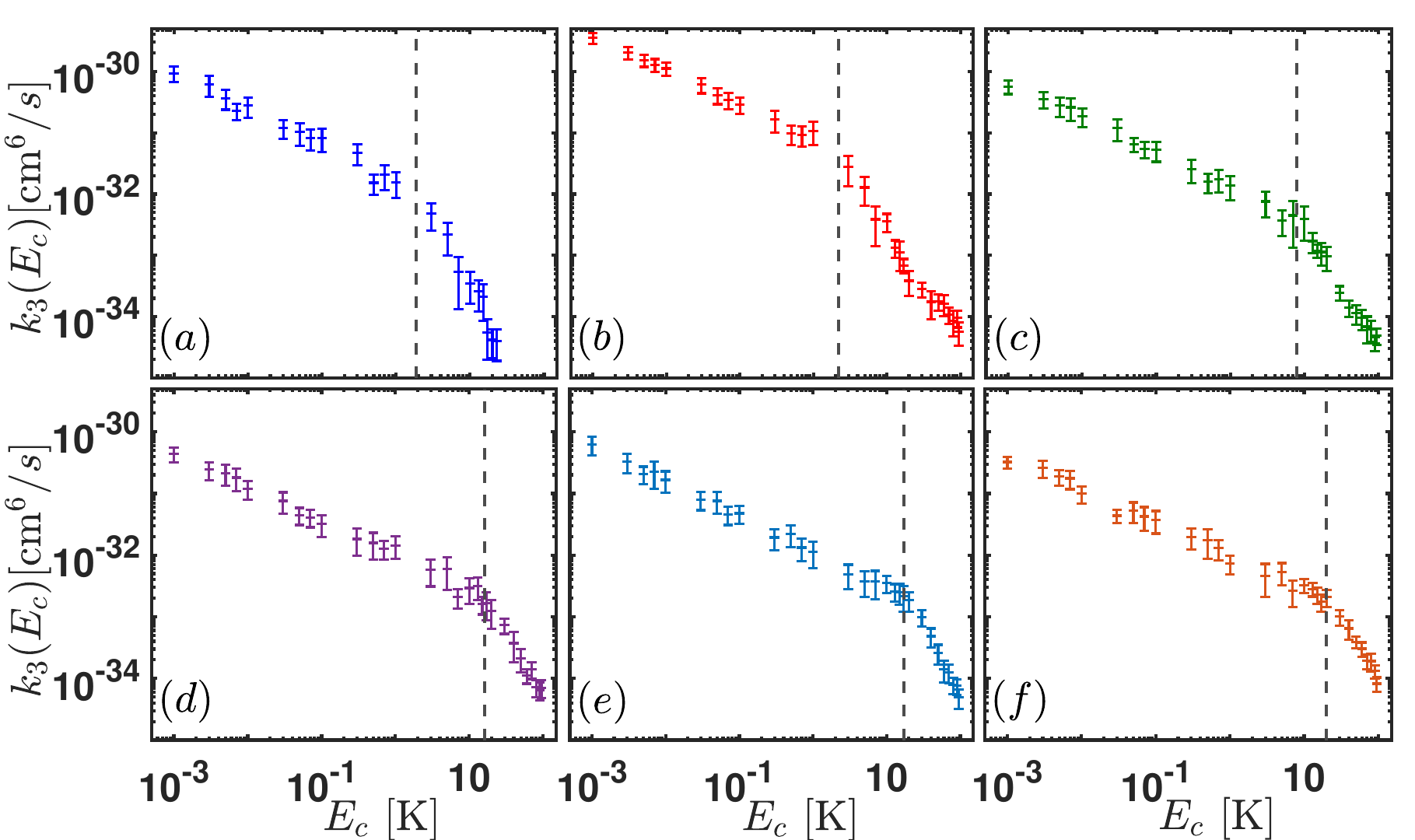}
			\caption{\label{fig:6atomk3}Three-body recombination rate of formation of six different X-He vdW molecules as a function of collision energy $E_c$, plotted on a log-log scale. The color code for panels $(a)$ to $(f)$ is same as that in \cref{fig:6atompots} where panel $(a)$ shows the rate for Na + He +He, and panel $(f)$ shows the rate related to N + He + He recombination. Each black dashed line indicates the relevant dissociation energy $D_e$. }
		\end{center}
	\end{figure*}

	The energy-dependent three-body recombination rates, $k_3(E_c)$, for the six considered cases are illustrated in \cref{fig:6atomk3}. It is quite remarkable that, despite the drastic differences in the properties of X atoms and parameters of X-He interaction potentials, the recombination rates are of the same order of magnitude.
	Moreover, it is noticed that the energy-dependent three-body recombination rate shows the same trend as a function of the collision energy, independently of the X atom under consideration. In particular, we identify two power-law behaviors (linear in the log-log scale) connected at the dissociation energy, $D_e$, represented as the black dashed line in each of the panels of \cref{fig:6atomk3}. Indeed, $D_e$ acts as the threshold energy for two distinct regimes: the low-energy regime, where $E_c<D_e$; and the high-energy regime, where $E_c>D_e$. 
	
	The data displayed in \cref{fig:6atomk3} shows that even though in both regimes, the dependence of $k_3$ on $E_c$ follows a power-law, the energy-dependence for the high-energy domain is much steeper than for the low-energy one. In our view, this behavior is related to the interplay between the role of the long-range tail of the X-He potential and its short-range region, in the formation of vdW molecules at different energies. In other words, the formation of vdW molecules at low energies is mainly a consequence of the X-He interaction potential's long-range tail, but this is not the case for high-energy collisions.
	
\begin{figure}[b]
		\begin{center}
			\includegraphics[scale=0.4]{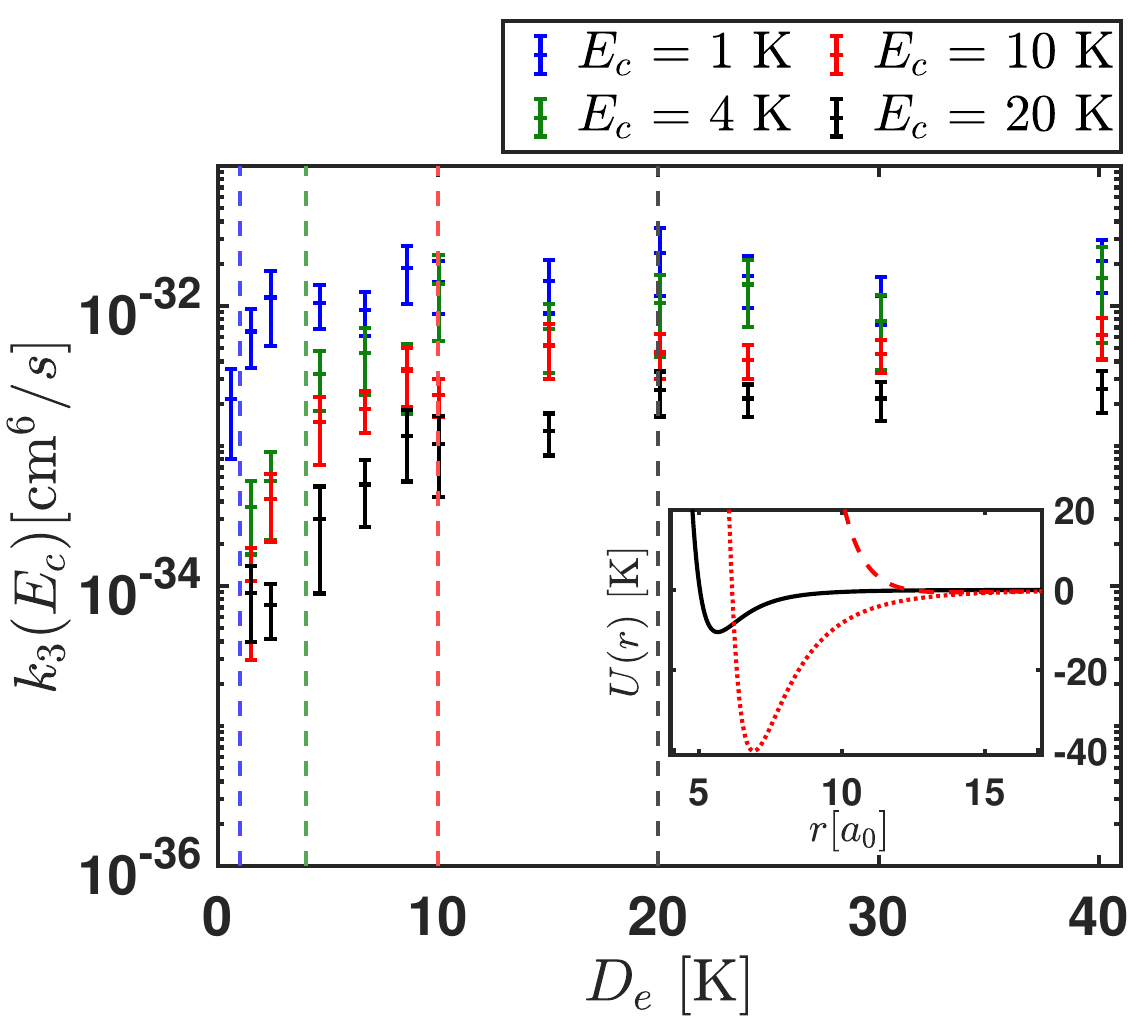}
			\caption{\label{fig:k3TiHec12int}Semi-logarithmic plots of energy-dependent three-body recombination rates for X-RG interaction being a LJ potential with constant $C_6 = 27.6$ [a.u.] and different $C_{12}\in\{1.5, 2, 2.5, 3, 4,  6, 7, 9,  13, 25, 40,  100\}\times 10^{6}$ [a.u.], for different collision energies $E_c\in\{1,4,10,20\}$ K. The Dashed lines indicate collision energies of the related color. The inset shows He-He potential (black curve) together with the deepest (red solid curve with $D_e = 40.13$ K) and the shallowest (red dashed curve with $D_e = 0.6$ K) LJ potentials.}
		\end{center}
	\end{figure}

	To check the validity of this statement, we have computed the energy-dependent three-body recombination rate at different collision energies by varying the short-range part of the interaction potential while keeping $C_6$ constant, and the results are shown in \Cref{fig:k3TiHec12int}. In this figure, we observe that only when $E_c>D_e$ the three-body recombination rate shows a variation from its nearly constant value at $E_c<D_e$. Therefore, the precise details of short-range X-He interaction (properties of potential well) only matter at high collision energies, where the three-body recombination rate starts to show a steep behavior. 
	
	The power-law behavior at low-energy collisions is expected in the virtue of the classical nature of the collisions \cite{Perez-Rios2014,Perez-Rios2020}, as it is explained below.

	\subsubsection*{Low-energy regime}
	Based on our results for the energy-dependent three-body recombination rates of X-He formation, we conclude that the energy-dependence of recombination rate in the low-energy regime depends chiefly on the dominant long-range $1/r^6$ interaction. To study this dependence, we apply a classical capture model following the pioneering ideas of Langevin for ion-neutral reactions \cite{Langevin1905}. In this framework, every trajectory with an impact parameter below some threshold value $\tilde{b}$ leads with unit probability to a reaction event. $\tilde{b}$, is given by the largest partial wave for which the height of the centrifugal barrier is equal to the collision energy. 
	For neutral-neutral interactions, the effective long-range potential reads as (in atomic units)	
	
	\begin{equation}\label{eq:Lang}
		U_\mathrm{eff}(r) = -\frac{C_6}{r^6} + \frac{\ell(\ell+1)}{2\mu_0r^2} ~.
	\end{equation}
	The second term in \cref{eq:Lang} is the centrifugal barrier with $\mu_0$ being the two-body reduced mass and $\ell$ the angular momentum quantum number or partial wave. The potential $U_\mathrm{eff}(r)$ shows a maximum at 
	
	\begin{equation}
		r_0 = \left[\frac{6\mu_0 C_6}{\ell(\ell+1)}\right]^{1/4}
	\end{equation}
	Classically, a reaction occurs if and only if $E_c \ge U_\mathrm{eff}(r_0)$. Now, to find the  critical impact parameter $\tilde{b}$ which is assigned to $E_c = U_\mathrm{eff}(r_0)$, we may use the relation between angular momentum quantum number $\ell$, the collision energy, and the impact parameter \cite{Levine2005,Perez-Rios2020}, i.e.,		
	\begin{equation}\label{eq:LPB}
		\ell(\ell+1) = 2\mu_0 \tilde{b}^2 E_c ~.
	\end{equation} 
	Substituting $\ell(\ell+1)$ obtained from  $E_c = U_\mathrm{eff}(r_0)$ into \cref{eq:LPB} yields
	
	\begin{equation}
		\tilde{b} = \sqrt{\frac{2}{3}} \left(\frac{2C_6}{E_c}\right)^{1/6} ~.
	\end{equation}
	
	Applying this model to both X-RG and RG-RG interactions and keeping in mind that 6D impact parameter $b$ is a combination of the 3D impact parameters associated with the Jacobi coordinates $\vec{\rho_1}$ and $\vec{\rho_2}$, we expect the same power-law for  $b_{max}$ (introduced in \cref{subsec:CTBR}), i.e, 
	\begin{equation}\label{eq:b6dL}
		b_{max} \propto E_c^{-1/6} ~.
	\end{equation}
	Therefore, in virtue of \cref{eq:b6dL} and the Langevin assumption that $\mathcal{P}(E_c,b>b_{max}) = 0$ and $\mathcal{P}(E_c,b\le b_{max}) = 1$, from \cref{eq:sigma} we obtain the low-energy power-law for the three-body recombination cross section as $	\sigma_{rec}(E_c) \propto E_c^{-5/6}$, and hence, the three-body recombination rate as
	
	\begin{equation}\label{eq:k3L}
		k_3(E_c) \propto E_c^{-1/3}  ~,
	\end{equation}
	which, as expected, it is consistent with the classical threshold law that has been found for low-energy collisions in Ref.~[\onlinecite{Perez-Rios2014}].
	
	\begin{figure}[ht]
		\begin{center}
			\includegraphics[scale=0.5]{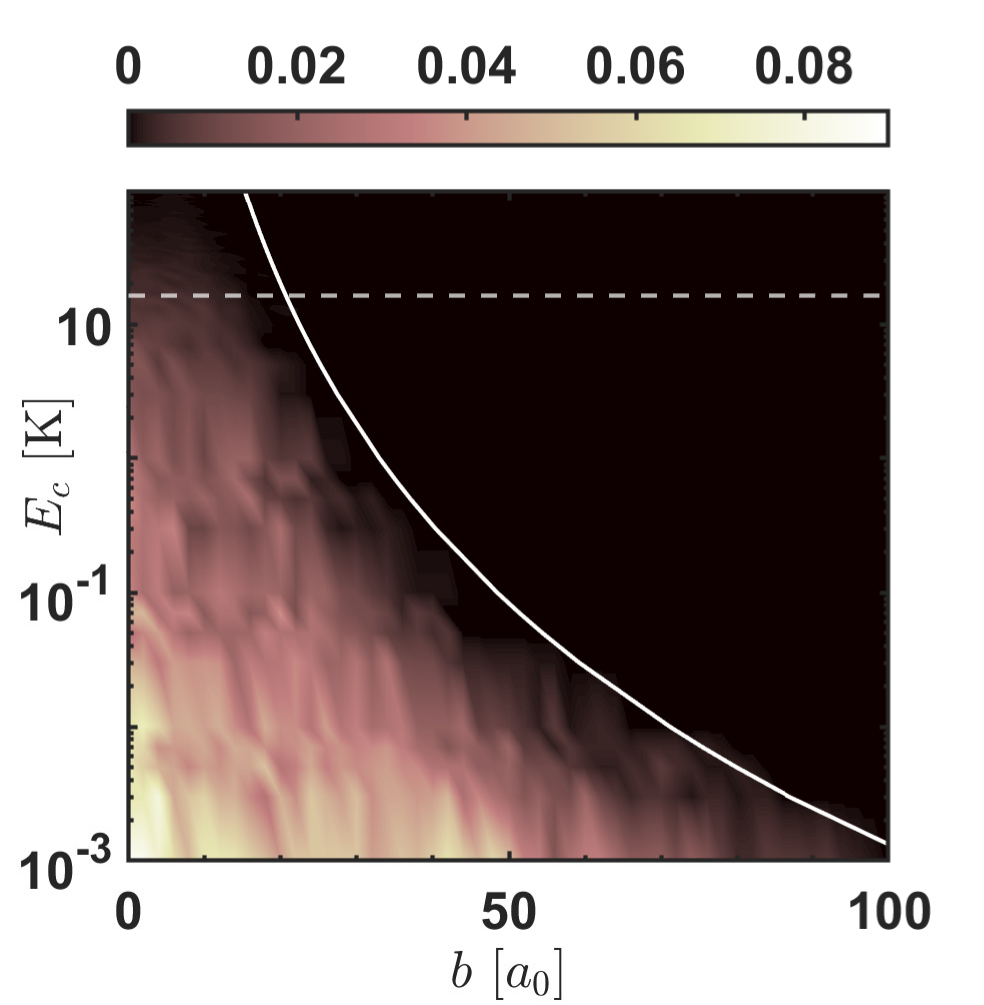}
			\caption{\label{fig:opacity_cplot}Opacity function $\mathcal{P}(E_c,b)$ of formation of As-He vdW molecules in As + He + He collisions, projected onto the $E_c-b$ plane with semi-logarithmic scale. The white curve shows $b \propto E_c^{-1/6}$, and the white dashed line indicates $D_e$.}
		\end{center}
	\end{figure} 
	Let us now examine our findings via an example, namely, As + He + He interaction. \cref{fig:opacity_cplot} shows the opacity function $\mathcal{P}(E_c,b)$ for the formation of As-He due to a three-body recombination in terms of collision energy $E_c$ and 6D impact parameter $b$. The white dashed line represents the collision energy equal to the dissociation energy, $E_c=D_e$, and the white curve indicates the $b \propto E_c^{-1/6}$. As expected, the opacity function has its maximum at $b = 0$ and $E_c = 10^{-3}$ K (the lowest illustrated energy). By increasing the impact parameter, the opacity function along each line at constant collision energy gradually decreases and eventually vanishes at $b = b_{max}$. 

	The white curve ($b \propto E_c^{-1/6}$) in \cref{fig:opacity_cplot}, reasonably resembles the loci of $b_{max}$ in the low-energy regime. However, in higher energies this loci deviates from the white curve and for collision energies $\gtrsim D_e$ (top left corner) does not obey the same power-low any more. Consequently, based on \cref{eq:k3L} we can now explain the observed trend of the recombination rates displayed in panel $(d)$ of \cref{fig:6atomk3}.
	While the energy-dependence of $k_3$ on the collision energies below $10^{-2}$ K can be conveniently explained by the adopted classical capture model, this model can not provide the correct power-law for the high energies above dissociation energy $D_e \approx 16$ K, where $b_{max}$ varies much faster than $E_c^{-1/6}$. In the intermediate regime connecting these two limits, the dependence of $k_3$ on the collision energies gradually deviates from the initial relation given by \cref{eq:k3L} and is closer to $k_3(E_c) \propto E_c^{-1/2}$. 
	
	In addition, it is worth mentioning that including the three-body interaction term of the X-He-He potential energy surface will lead to a deviation from the derived power-law behavior for $b_{max}$ and $k_3(E_c)$.
	 
	\subsubsection*{High-energy regime}
\begin{figure}
		\begin{center}
			\includegraphics[scale=0.43]{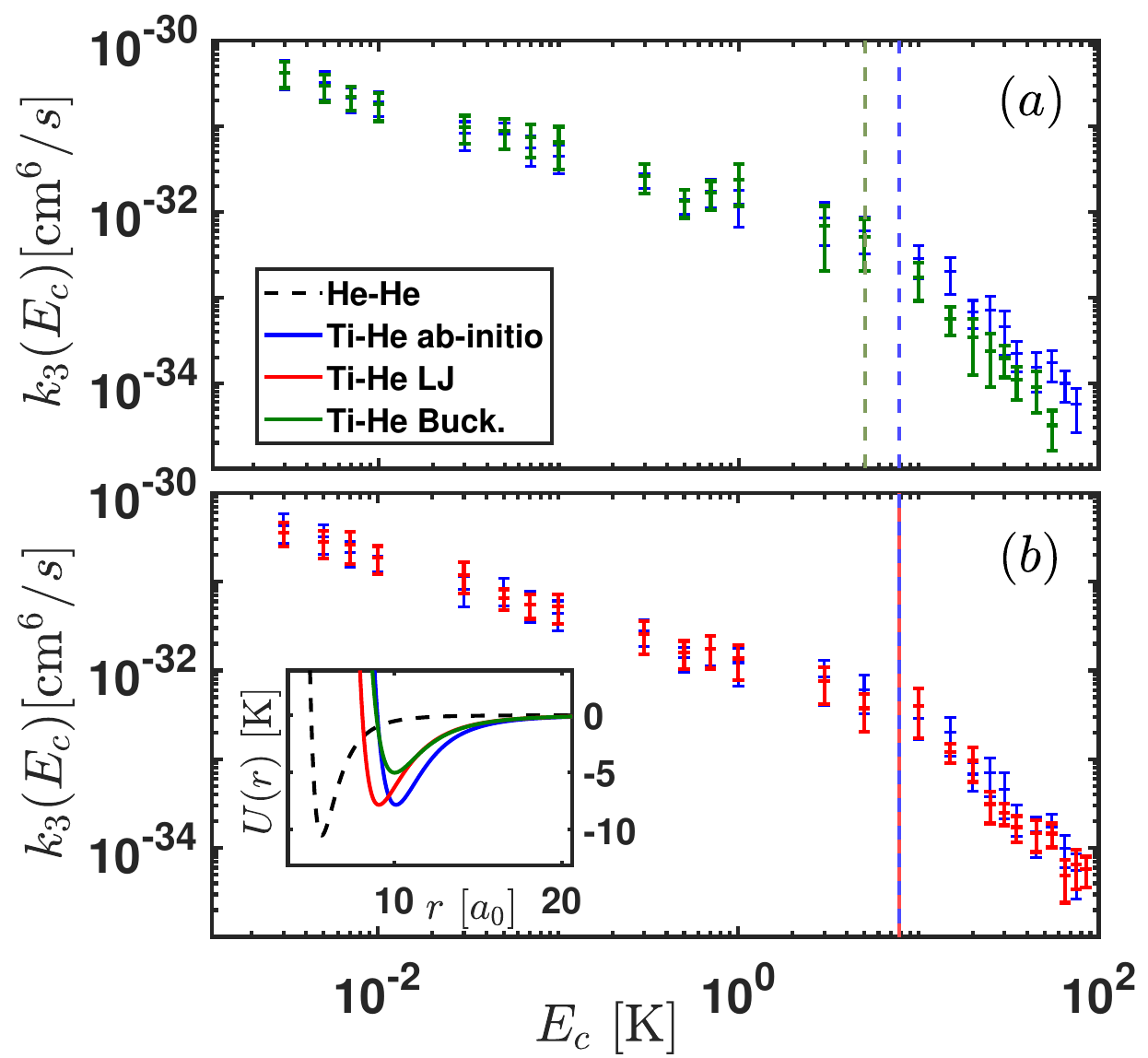}
			\caption{\label{fig:K3(E)_TiHe}Comparison of the three-body recombination rate leading to the formation of Ti-He vdW molecules as a function of collision energy $E_c$ (log-log plot) between {\it ab initio} potential and $(a)$ LJ, and $(b)$ Buckingham (Buck.) potentials shown in the inset. The blue, red, and green dashed lines indicate dissociation energies related to {\it ab initio}, LJ, and Buckingham potentials, respectively. }
		\end{center}
	\end{figure}

    To explore the effect of short-range details of potential on the formation of vdW molecules at higher collision energies, without loss of generality, we focus on the energy-dependent three-body recombination rate for Ti + He + He. The CT calculations are performed by means of three different potentials for the Ti-He interaction, namely, {\it ab initio} potential form Refs.~[\onlinecite{Krems2004,Krems2005,Quiros2017}], Buckingham potential \cite{Buckingham1938} with the form $U(r) = C_1\exp(-C_2 r) - C_6/r^6 $, and the LJ potential used in previous calculations. The results are displayed in \cref{fig:K3(E)_TiHe}. The $C_6$ dispersion coefficient in the LJ and Buckingham potentials is derived from {\it ab initio} calculations \cite{Krems2005,Quiros2017}. 
	
    As expected from our previous discussion, despite the non-negligible differences in the shape of the short-range interaction potential in the short-range region (compare red (LJ), blue ({\it ab initio}), and green (Buckingham) curves in the inset of \cref{fig:K3(E)_TiHe}), the three-body recombination rates $k_3(E_c)$ at collision energies below the dissociation energy ($E_c<D_e$) are the same, which confirms our observation that low-energy collisions are dominated by the long-range tail of the X-He potential. 
    
	In contrast, proceeding to the high-energy regime, we spot two distinct behaviors. First, the three-body recombination rate related to the shallower X-RG potential (smaller $D_e$) shows a slightly steeper energy dependence and accordingly, the power-law is different (see panel $(a)$ in \cref{fig:K3(E)_TiHe}). Second, the energy-dependence of the three-body recombination rate do not depend on the equilibrium distance of the X-He molecule as long as the dissociation energy is the same (see panel $(b)$ in \cref{fig:K3(E)_TiHe}). In other words, the dissociation energy seems to be the most relevant short-range parameter of the two-body potential affecting the recombination rate.  
	Therefore, the inclusion of non-additive interactions on the X-RG-RG potential energy surface may lead to a slightly different trend.
	
	It is important to note that the long-range tail of the {\it ab initio} potential contains higher order terms proportional to $1/r^8$, $1/r^{10}$, $\dots$ coming from the spherical multipole moment expansion of the involved electronic clouds. However, the three-body recombination rates obtained for both LJ and {\it ab initio} potentials are identical in the whole energy regime as well as for Buckingham and {\it ab initio} potentials in the low-energy regime. Therefore, our results strongly suggest that the effect of long-range interaction on formation of vdW complexes is mainly through the $ 1/r^6$ term of the dispersion potential. 

	\subsection{Temperature-dependent three-body recombination rate for X-He-He systems}
	    
    \begin{figure*}
		\begin{center}
			\includegraphics[scale=0.5]{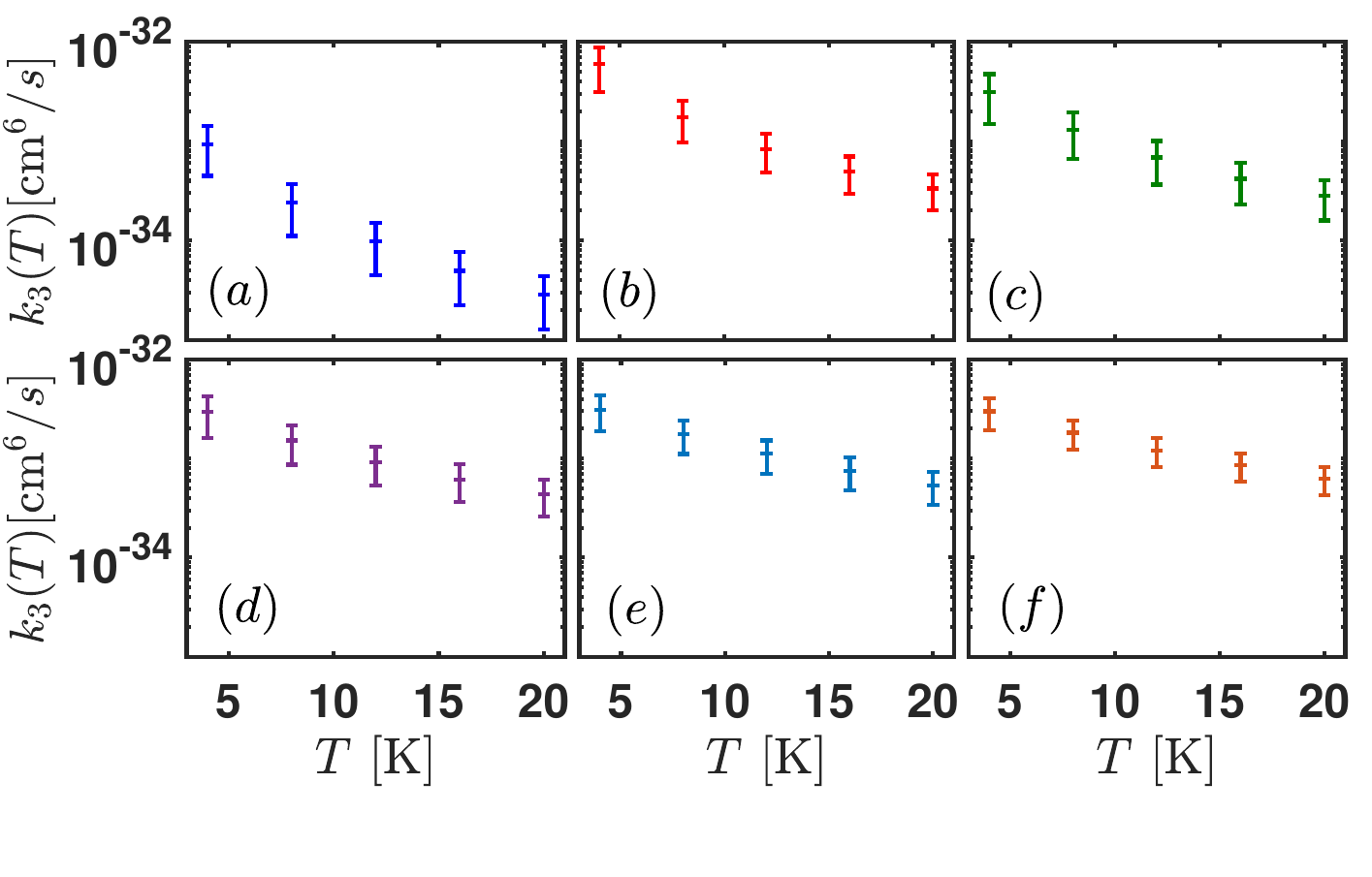}
			\caption{\label{fig:6atomk3T}Three-body recombination rate of formation of six different X-He vdW molecules as a function of temperatures $T\in\{4,8,12,16,20\}$ K (semi-log plots). The color code for panels $(a)$ to $(f)$ is same as that in \cref{fig:6atompots} where panel $(a)$ shows the rate for Na + He +He, and panel $(f)$ shows the rate related to N + He + He recombination.}
		\end{center}
	\end{figure*}
	
	In the final part of our discussion, we focus on the production of vdW molecules 
	in buffer gas cells. In particular, we investigate the thermal averaged three-body recombination rate as a mechanism for the formation of X-He vdW molecules at temperatures $4 \leq T \leq 20$ K. The thermal average of the three-body recombination rate is obtained via integrating the energy-dependent three-body recombination rate \cref{eq:k3} over the appropriate three-body Maxwell-Boltzmann distribution of collision energies, yielding
	
	\begin{equation}\label{eq:MBa}
		 k_3(T) =  \frac{1}{2(k_BT)^3} \int_{0}^{\infty}k_3(E_c) E_c^2 e^{-E_c/(k_BT)} dE_c ~,
	\end{equation}
    where $k_B$ is the Boltzmann constant. The results obtained by performing the thermal average~\eqref{eq:MBa} for six different X + He + He reactive collisions for $4 \leq T \leq 20$ K, are shown in \cref{fig:6atomk3T}. Comparing the data in panels $(d)$, $(e)$, and $(f)$ of \cref{fig:6atomk3T}, it is noticed that the trend and the magnitude of the three-body recombination rate is very similar for all three considered pnictogens As, P, and N. Similarly, the temperature-dependent three-body recombination rate displays the same tendency for alkali-metals Li and Na (panels  $(a)$ and $(b)$).

    Finally, we notice that the three-body recombination rate $k_3(T)$ for all the X atoms considered shows nearly the same order of magnitude, except for Na. This is due to the rapid decrease of the Na-He recombination rate $k_3(E_c)$ at relatively low collision energies in comparison to the rest of X-He complexes (compare panel $(a)$ with other panels of \cref{fig:6atomk3}). However, it is still fascinating that vdW molecules containing X atoms from totally different groups of periodic table show a similar three-body recombination rate within the range of typical temperatures in buffer gas cells. Indeed, in virtue of the observation of Li-He \cite{Tariq2013}, Ag-He \cite{Brahms2008}, and Ti-He \cite{Quiros2017}, it should be possible to observe any X-He molecule in a buffer gas cell under the proper conditions.   
    
    \section{Reliability of classical trajectory calculations at low temperatures}\label{sec:rel}
    \begin{table}[b]
		\caption{\label{tab:table1}The largest classically allowed partial wave $\ell_\mathrm{max}$ (see \cref{lmax}) contributing to the scattering observables for collision energy $E_c$.}
		\begin{ruledtabular}
			\begin{tabular}{rccc}
				&\multicolumn{3}{c}{$E_c ~[\mathrm{K}]$}\\ 
				X-RG & $100$ & $10$ & $1$  \\ 
				\hline 
				Li-He\footnotemark[1] & $15$ & $7$ & $3$  \\
				-Ne\footnotemark[2] & $24$ & $11$ & $5$   \\
				-Ar\footnotemark[3] & $32$ & $15$ & $7$  \\
				-Kr\footnotemark[3] & $36$ & $16$ & $7$  \\
				-Xe\footnotemark[3] & $39$ & $18$ & $8$  \\
				Na-He\footnotemark[1] & $17$ & $8$ & $3$  \\
				-Ne\footnotemark[2] & $35$ & $16$ & $7$  \\
				-Ar\footnotemark[3] & $51$ & $24$ & $11$ \\
				-Kr\footnotemark[3] & $61$ & $28$ & $13$  \\
				-Xe\footnotemark[3] & $68$ & $31$ & $14$  \\
				N-He\footnotemark[4] & $13$ & $6$ & $2$ \\
				-Ar\footnotemark[5] & $33$ & $15$ & $7$  \\
				-Kr\footnotemark[5] & $39$ & $18$ & $8$  \\
				P-He\footnotemark[4] & $16$ & $7$ & $3$  \\
				As-He\footnotemark[4] & $17$ & $8$ & $3$  \\
				Ti-He\footnotemark[6] & $18$ & $8$ & $4$  \\
			\end{tabular}
		\end{ruledtabular}
		\footnotetext[1]{vdW coefficient $C_6$ is taken from Ref.~[\onlinecite{Kleinekathofer1999}].}
		\footnotetext[2]{vdW  coefficient $C_6$ is taken from  Ref.~[\onlinecite{Tao2010}].}
		\footnotetext[3]{vdW  coefficient $C_6$ is taken from Ref.~[\onlinecite{Tangt1976}].}
		\footnotetext[4]{vdW  coefficient $C_6$ is taken from  Ref.~[\onlinecite{Partridge2001}].}
		\footnotetext[5]{vdW  coefficient $C_6$ is taken from  Ref.~[\onlinecite{Cambi1991}].}
		\footnotetext[6]{vdW  coefficient $C_6$ is taken from Refs.~[\onlinecite{Krems2005,Quiros2017}].}
	\end{table}
	
	The reliability of classical trajectory calculations for scattering observables depends on the collision energy at which the system is studied. In particular, the lower the collision energies, the more quantum mechanical effects become prominent. In general, the importance of quantum mechanical effects on a system is related to the number of partial waves contributing to the scattering observables. Classically, the largest number of the allowed partial wave is calculated by setting the centrifugal barrier equal to the collision energy $E_c$ \cite{Perez-Rios2020}. For the systems under consideration in this work, the two-body X-RG long-range interaction (i.e., $-C_6/r_\mathrm{X-RG}^6$) dominates the potential interaction. Thus, we have
	\begin{equation}\label{lmax}
		\ell_\mathrm{max} =  \sqrt{6\mu_0} C_6^{\frac{1}{6}}  \left( \frac{E_c}{2}  \right)^{\frac{1}{3}}, 
	\end{equation}
	with $\mu_0$ being the two-body reduced mass.
	
	As an example, $\ell_{max}$ values for the X-He pairs are listed in \cref{tab:table1}. Note that in the case of three-body collisions, the total angular momentum will be affected by the He-He interaction, and the expected number of partial waves may increase compared with the one obtained from the X-RG long-range interaction potential. Moreover, based on \cref{lmax}, at a given collision energy, heavier RG atoms show larger $\ell_{max}$. This can be understood, considering that the heavier the system is, the closer to the classical realm it is. In this case, it is also related to the fact that heavier RG atoms show larger static polarizability and hence a larger $C_6$ (in general).
	
	Therefore, due to the relatively large number of partial waves, we believe that the CT calculation presented in \cref{sec:CT} is a reasonable approach to study the formation of vdW molecules even at energies near 4 K. For a more detailed comparison between the quantum and classical results obtained by hyperspherical CT method in three-body collisions see Ref.~[\onlinecite{Perez-Rios2014}].
	
	\section{Conclusions and prospects}\label{sec:conclusion}
	In summary, we have shown, via a classical trajectory method introduced in Ref.~[\onlinecite{Perez-Rios2014}], that van der Waals molecules can be formed through direct three-body recombination. In particular, we have investigated the energy-dependence and temperature-dependence of the three-body recombination rates for six vdW complexes containing atoms with totally different chemical properties. As a result, we found the X-RG molecule's dissociation energy is the determinant parameter to differentiate between the low and high energy regimes. At low energies, the formation rate of vdW molecules is relatively insensitive to the short-range interaction and is dominated by the long-range tail of the potential, i.e.,  $-1/r^6$. This regime is further explored by a classical capture model, \`a la Langevin. Conversely, at higher collision energies, the short-range part of the potential plays an important role. 

	However, the most exciting result is that the three-body recombination rate for the formation of vdW molecules is of the same order of magnitude, independently of the chemical properties of the atom colliding with the remaining two rare gas atoms. In other words, the formation rate of X-RG vdW molecules is almost independent of the chemical properties of X atom. Indeed, we have shown that CT calculations are reliable for relevant temperatures in buffer gas cells (1K~$\lesssim T \lesssim 10$ K) based on the number of contributing partial waves to different scattering observables. Therefore, it should be possible to create and study X-RG vdW molecules in buffer gas cells, where some of them have been already observed and studied. 
	Nevertheless, it is necessary to correctly identify the molecular dissociation processes to understand vdW molecules in equilibrium conditions, which can be considered as an extension of this work in the near future. Last but not least, in our view, a complete understanding of the formation process of vdW molecules, will become a cornerstone of the physics of vdW complexes' aggregation. 
	
   \begin{acknowledgments}
   We thank Dr. Timur V. Tscherbul for sharing with us the {\it ab initio} potential of Ti-He. 
   \end{acknowledgments}

	\bibliographystyle{rsc}
	\bibliography{Formation_vdW.bib}
\end{document}